 \documentclass[12pt]{article}
   \usepackage{amssymb}

 \newcommand{\beq}[1]{\begin{equation}\label{#1}}
 \newcommand{\eeq}{\end{equation}}
 \newcommand{\bear}[1]{\begin{eqnarray}\label{#1}}
 \newcommand{\ear}{\end{eqnarray}}
 \newcommand{\nn}{\nonumber}

 \renewcommand{\theequation}{\arabic{section}.\arabic{equation}}
 \catcode`\@=11 \@addtoreset{equation}{section}\catcode`\@=12

 \newcommand{\np}{ {\newpage } }
 \newcommand{\R}{ \mathbb{R} }
 
 \newcommand{\e}{ \mbox{\rm e} }
 \newcommand{\eps}{ \varepsilon }
 \newcommand{\p}{\partial}
 \newcommand{\btd}{\bigtriangledown}
 \newcommand{\btu}{\bigtriangleup}

 \newcommand{\fnm}{\footnotemark}
 \newcommand{\fnt}{\footnotetext}

\begin{document}

 \begin{center} \large \bf

  $S$-brane  solutions with
  (anti-)self-dual parallel charge density form on a Ricci-flat
   submanifold

 \end{center}

 \vspace{0.63truecm}

 \bigskip

 \begin{center}

 \normalsize

 \bf  H. Dehnen\fnm[1]\fnt[1]{Heinz.Dehnen@uni-konstanz.de},

  \bigskip

  \it  Universit\"at Konstanz, Fachbereich Physik,
       Fach  M 677, D-78457, Konstanz

 \bigskip

 \bf   V.D. Ivashchuk\fnm[2]\fnt[2]{rusgs@phys.msu.ru},
       V.N. Melnikov\fnm[3]\fnt[3]{melnikov@phys.msu.ru}

 \bigskip

 \it  Center for Gravitation and Fundamental Metrology,
  VNIIMS and  Institute of Gravitation and Cosmology,
  Peoples' Friendship University of Russia,
  46 Ozernaya Street, Moscow, 119361, Russia

  \bigskip

 \end{center}

  \begin{abstract}

  A $D$-dimensional cosmological model
  with several scalar fields and  antisymmetric $(p+2)$-form is considered.
  For dimensions $D = 4m+1 = 5, 9, 13, ...$ and
  $p = 2m-1 = 1, 3, 5, ...$ we obtain a family of new cosmological type
  solutions with $4m$-dimensional oriented Ricci-flat  submanifold $N$ of
  Euclidean signature.  These solutions  are characterized by a self-dual or
  anti-self-dual parallel  charge density form $Q$ of rank $2m$ defined
  on $N$. The (sub)manifold $N$ may be chosen to be K$\ddot{a}$hler, or 
  hyper-K$\ddot{a}$hler one, or  8-dimensional manifold of $Spin(7)$ 
  holonomy.  The  generalization of solutions to a chain of extra 
  (marginal) Ricci-flat factor-spaces is also presented.  Solutions with 
  accelerated expansion of extra factor-spaces are singled out.  Certain 
  examples of new solutions for IIA supergravity and for a chain of 
  $B_D$-models in dimensions $D = 14, 15,...$ are considered.

 \end{abstract}



 \np

 \section{\bf Introduction}
 \setcounter{equation}{0}

   This paper is devoted to generalization of composite electric S-brane
   solutions with maximal number of branes obtained earlier in   \cite{IS,DIM,IMS}.
  (For $S$-brane solutions see \cite{DHHS}-\cite{HLPS} and refs.  
  therein.) These solutions exist in gravitational models in dimensions $D 
   = 4m+1 = 5, 9, 13, ...$ and containing $(p + 2)$-form with $p = 2m-1 = 
   1, 3, 5, ...$.  An interesting feature of these "maximal"  solutions is  
   the linear relations between  charge densities \cite{IS}.

    Namely, it was shown in \cite{IS} that electric $S$-brane solutions with maximal number
   of branes in $5$-dimensional model   with $3$-form  and scalar field exists
    when the charge densities
   of six electric branes obey the following relations
    \bear{b1}
    Q_{12} = \mp Q_{34}, \quad Q_{13} = \pm Q_{24}, \quad
     Q_{14} = \mp Q_{23}.
     \ear
   or, equivalently,
    \beq{b2}
    Q_{i j} =  \pm \frac{1}{2} \eps _{i j k l}
      Q^{k l} = \pm (* Q)_{i j}.
     \eeq
   Here $Q_{i j} = - Q_{j i}$.
   When all $Q_{i j}$ ($i \neq j$) are non-zero the configuration from
   (\ref{b2}) is the only possible one that follows just from
   non-diagonal part of Hilbert-Einstein equations.

    Analogous  (anti)-self-duality relations were used in constructing
    exact solutions in   dimensions $D = 4m+1 = 5, 9, 13, ...$   \cite{IS,DIM,IMS}.

    In the case when the scalar field is absent we are led in \cite{IMS}
    to a solution for $D= 5$ gravity with $3$-form and found the absence
    of oscillating behaviour in this model when cosmological electric
    S-brane solutions with diagonal metric are considered. This behaviour
    corresponds to the case of a "frozen" point in a billiard when we
    approach to the singularity (for billiard approach with branes see
    \cite{IMb,DHN}).

    It should be noted that the relations between charge densities 
    of $D$-branes
   play an important role  in the string theory, see, for example,
   formulas between charge densities of electric and magnetic branes
   in ref. \cite{Polchinsky}.  It was  shown  that such relations
   in a general case may be described by mathematical  $K$-theory, see
   \cite{MM,Witten,Gukov} and references therein. But these
   relations are of topological origin, while the relations
   proposed in our papers are of dynamical origin: they
   appear as solutions of quadratic constraints on charge densities following
   from non-diagonal part of Hilbert-Einstein equations.

   Here, in Section 4,  we obtain  new cosmological type solutions
  with an oriented  Ricci-flat factor-space $N$ of dimension $4m$
  and self-dual or anti-self-dual
  non-zero parallel (i.e. covariantly constant)  "charge density" form $Q$
  of rank $2m$ defined on $N$.  The examples of Ricci-flat Riemannian
  manifolds of dimension $4m$ equipped with (anti-)self-dual parallel
  $2m$-form contain:  K$\ddot{a}$hler manifolds of holonomy group
  $SU(2m)$, hyper-K$\ddot{a}$hler  manifolds with holonomy group $Sp(m)$
  and 8-dimensional Ricci-flat manifold of  $Spin(7)$ holonomy \cite{Besse}.
  In Section 5 we  generalize the solutions from  Section 4 to the case
  when a chain of extra  Ricci-flat factor-spaces is added into
  consideration.  Here we  obtain a family of new solutions with
  accelerated expansion of extra factor-spaces. In Section 6 we present
   examples of new solutions for IIA supergravity and for a chain
  of the so-called $B_D$-models from \cite{IMJ} in dimensions $D =  14,
  15,...$. In Appendix we give an explicit derivation of exact solutions
  from Section 5.

 \section{\bf The model}
 \setcounter{equation}{0}

As in \cite{DIM} we consider the model governed by the action
   \beq{2.1i}
    S =
       \int_{M} d^{D}z \sqrt{|g|} \left[ {R}[g] -
    m_{ab}  g^{MN} \partial_{M} \varphi^a \partial_{N} \varphi^b
    -  \frac{1}{q!} \exp( 2 \lambda_a \varphi^a ) F^2 \right],
   \eeq
 where $g = g_{MN} dz^{M} \otimes dz^{N}$ is the metric,
 $\vec{\varphi} = (\varphi^a)$   is a set  (vector) of scalar fields,
 $a = 1, \ldots, l$;  $\vec{\lambda} = (\lambda_a) \in  \R^l$ is a
 constant vector (set) of dilatonic couplings,
 $(m_{ab})$ is a symmetric non-degenerate $l \times l$ matrix,
   \beq{2.2i}
   F =  dA =
   \frac{1}{q!} F_{M_1 \ldots M_{q}}
   dz^{M_1} \wedge \ldots \wedge dz^{M_{q}},
   \eeq
 is a $q$-form, $q =  p +2 \geq 1$, on a $D$-dimensional manifold
 $M$.

 In (\ref{2.1i}) we denote $|g| = |\det (g_{MN})|$, and
   \beq{2.3i}
   F^2 =
         F_{M_1 \ldots M_{q}} F_{N_1 \ldots N_{q}}
         g^{M_1 N_1} \ldots g^{M_{q} N_{q}}.
   \eeq

 The equations of motion corresponding to  (\ref{2.1i}) are
   \bear{2.4i}
   R^M_N - \frac{1}{2} \delta^M_N R  =   T^M_N,
   \\
   \label{2.5i}
   {\btu}[g] \varphi^a -  \frac{\lambda^a}{q!}
    e^{2 \lambda (\varphi)} F^2 = 0,
   \\
   \label{2.6i}
   \nabla_{M_1}[g] (e^{2 \lambda (\varphi)}
    F^{M_1 \ldots M_{q}})  =  0.
   \ear

 Here, and in what follows $\lambda (\varphi)= \lambda_a
 \varphi^a$.
 In (\ref{2.5i}) and (\ref{2.6i}), ${\btu}[g]$ and ${\btd}[g]$ are
 Laplace-Beltrami and covariant derivative operators corresponding
 to  $g$. Equations (\ref{2.4i}), (\ref{2.5i}) and (\ref{2.6i})
 are,  respectively, the multidimensional Einstein-Hilbert
 equations, the "Klein-Fock-Gordon" equation for the scalar field
 and the "Maxwell" equations for the $q$-form.

 The source terms in (\ref{2.4i}) can be split up as
   \bear{2.7i}
   T^M_N =   T^M_N [\varphi,g]
   + e^{2 \lambda (\varphi)} T^M_N[F,g],
   \ear
   with
   \bear{2.8i}
   T^M_N[\varphi,g] =
   m_{ab}[ g^{MQ} \p_{Q} \varphi^a \p_{N} \varphi^b -
   \frac{1}{2} \delta^M_N g^{PQ} \p_{P} \varphi^a \p_{Q} \varphi^b ],
   \\
   T^M_N [F,g] = \frac{1}{q!} \left[ - \frac{1}{2} \delta^M_N F^2
   + q  F_{N M_2 \ldots M_{q}} F^{M M_2 \ldots M_q}\right] ,
   \label{2.9i}
   \ear
 being the stress-energy tensor of the scalar field and $q$-form,
 respectively.

 \section{\bf S-brane solutions with flat factor spaces}
 \setcounter{equation}{0}

    Now, we describe  $Sp$-brane solutions from \cite{DIM}
    in dimensions
      \beq{3.h}
         D= n+1 = 4m + 1= 5, 9, 13, \dots
      \eeq
    with
      \beq{3.p}
         p = 2m - 1= 1, 3, 5, \dots
      \eeq
    and non-exceptional dilatonic coupling
     \beq{lambda}
      \lambda ^2 \equiv  m^{ab} \lambda_a \lambda_b
      \neq  \frac{n}{4(n-1)} \equiv \lambda^2_0.
     \eeq
    Here, and in what follows the matrix $(m^{ab})$
      is inverse to $(m_{ab})$.

    These solutions are defined on  the manifold
   \beq{2.10g}
    M = (u_{-}, u_{+})  \times \R^{n}
   \eeq
    and have the following form
     \bear{4.pa}
      ds^2
      &=& we^{2n \phi(u)} du^2 + e^{2 \phi(u)}
     \sum _{i=1}^n (dy^i)^2 \\
     \label{4.pb}
     \varphi^a &=&
               - \frac{\lambda^a}{K} [ f(u) +  \lambda_b (C_2^b u + C_1^b)]
                +  C_2^a u + C_1^a, \\
     \label{4.pc}
     F &=& e^{2 f(u)} du \wedge Q,   \\
     \label{4.pq}
      Q &=&  \frac{1}{(p+1)!} Q_{i_0  \dots i_p}
                           dy^{i_0}  \wedge \dots  \wedge dy^{i_p}.
     \ear

     $w = \pm 1$, $\lambda^a = m^{ab} \lambda_b$,
     $C_2^a , C_1^a$ are integration constants,
     \beq{4.k}
          K \equiv  \lambda^2 - \frac{n}{4(n-1)} ,
      \eeq
     and $Q_{i_0  \dots i_p}$ are constant
     components of charge density form $Q$.
     $Q$ is a self-dual or anti-self-dual in a flat
     Euclidean space $\R^n$, i.e.
     \beq{3.eaa}
      Q_{i_0  \dots i_p} =
     \pm \frac{1}{(p+1)!} \eps _{i_0  \dots i_p j_0 \dots j_p}
      Q^{j_0 \dots j_p} = \pm (* Q)_{i_0  \dots i_p}.
     \eeq

     The function   $\phi (u)$ is defined as follows
     \beq{4.n}
       \phi = \frac{1}{2 (1- n) K} \left[ \lambda_a (C_2^a u + C_1^a) +
             f(u) \right],
       \eeq

     and the function   $f(u)$ is given by  relations
        \beq{4.i}
        f = - \ln \left[|z ||K Q^2|^{1/2} \right]
     \eeq
     with
     \bear{4.ja}
     z &=& \frac{1}{\sqrt{C}} \sinh \left[ (u-u_0) \sqrt{C} \right],
           \qquad K<0 , ~C>0; \\
                          \label{4.jb}
     &=& \frac{1}{\sqrt{-C}} \sin \left[ (u-u_0) \sqrt{-C} \right],
           \qquad K<0 , ~C<0; \\
                           \label{4.jc}
     &=& u-u_0, \qquad \qquad \qquad \qquad \qquad K<0, ~C=0; \\
                            \label{4.jd}
      &=& \frac{1}{\sqrt{C}} \cosh \left[ (u-u_0) \sqrt{C} \right],
     \qquad K>0 , ~C>0,
     \ear
     where
      \beq{4.q}
       Q^2 \equiv \frac{1}{(p+ 1)!}
       \sum_{i_0,  \dots, i_p} Q_{i_0  \dots i_p}^2 > 0,
      \eeq
     and
      \beq{4.o2}
      C = (\lambda_a C_2^a)^2 - K m_{ab}C_2^a C_2^b.
      \eeq

      The solution   presented above for $w =  -1$
      describes a collection of
      $k \leq C_{4m}^{2m}$ electric $Sp$-branes with
      non-zero charge densities $Q_{i_0  \dots i_p} \neq 0$,
      $i_0 <  \dots < i_p$, $p = 2m -1$.

      The case of one scalar field was  considered previously in \cite{IS}
      and in \cite{IMS} the solutions without scalar fields were
      studied.

 \section{\bf Generalization to Ricci-flat factor space}
 \setcounter{equation}{0}

    Here, we generalize the solution from the previous
    section to the case when the manifold
    (\ref{2.10g}) is replaced by the manifold
    \beq{3.10g}
    M = (u_{-}, u_{+})  \times N,
   \eeq
    where $N$ is $n$-dimensional oriented  manifold
    of dimension $n =4m$, $m =1,2, \dots$, equipped with the
    Ricci-flat metric $h = h_{ij}(y)dy^i \otimes dy^j$ of Euclidean
    signature.

    Let
    \beq{3.11}
      Q =  \frac{1}{(p+1)!} Q_{i_0  \dots i_p}(y)
              dy^{i_0}  \wedge \dots  \wedge dy^{i_p}
    \eeq
     be a non-zero form of rank $2m$ defined on the manifold $N$.

    The form $Q$ is supposed to be parallel, i.e.
    covariantly constant w.r.t. $h$
     \beq{3.12a}
      (i) \quad \btd[h] Q = 0,
     \eeq
   and also  self-dual or anti-self-dual one
     \beq{3.12b}
      (ii) \quad  Q = \pm * Q.
     \eeq
     Here $* = *[h]$ is the Hodge operator corresponding to
     the metric $h$.

      It follows from  (i)  that the form $Q$ is closed
      ($dQ = 0$) and 
      \beq{3.12ca}
        \quad Q^2 \equiv \frac{1}{(p+ 1)!}
       h^{i_0 j_0} \dots h^{i_p j_p}
       Q_{i_0  \dots i_p} Q_{j_0  \dots j_p}
        \eeq
      is  constant.

     Obviously,
        \beq{3.12c}
          Q^2 > 0,
        \eeq
 since $Q$ is non-zero and $h$ has Euclidean signature.

      For non-exceptional value of the dilatonic coupling
     (\ref{lambda}) the following solution on the
     manifold (\ref{3.10g}) is valid

     \bear{5.pa}
      ds^2
      &=& we^{2n \phi(u)} du^2 + e^{2 \phi(u)}
      h_{ij}(y)dy^i dy^j, \\
     \label{5.pb}
     \varphi^a &=&- \frac{\lambda^a}{K}
     [ f(u) +  \lambda_b (C_2^b u + C_1^b)]
                +  C_2^a u + C_1^a,   \\
     \label{5.pc}
     F &=& e^{2 f(u)} du \wedge Q,
     \ear
     with the $2m$-form  (\ref{3.11}) obeying
     (\ref{3.12a}), (\ref{3.12b}) and (\ref{3.12c}).

     The function  $\phi (u)$ is given by  (\ref{4.n}),
     where  $K$ is defined in  (\ref{4.k}),
      $C_2 , C_1$ are integration constants and
     the function   $f(u)$ is presented  by  relations
     (\ref{4.i})-(\ref{4.jd}) with  $Q^2$
     defined in  (\ref{3.12ca}). The integration constants
     $C$ and $C_2$ are related by the formula  (\ref{4.o2}).

     Now, we show that the metric with the line element (\ref{5.pa}),
     scalar fields  (\ref{5.pb}) and  $2m$-form (\ref{5.pc}) does
     satisfy  the field equations  (\ref{2.4i})-(\ref{2.6i}).

    For scalar fields $\varphi^a = \varphi^a(u)$ we get from   (\ref{2.5i})
    for Ricci-flat $h$  the same ordinary differential equations as in
    the flat case, so eqs. (\ref{2.5i}) are satisfied.

     The relation (\ref{3.12a})  reduces the verification of
     the  ``Maxwell'' equation  (\ref{2.6i}) just to the
     identity that holds for a flat-case solution from the previous
     section.

     Now, we verify the Hilbert-Einstein eqs. (\ref{2.4i}).

     First, we prove that
    \beq{5.d}
     T[F,g]_i^{~j} = 0,
    \eeq
   for all $i,j = 1, \dots, n$, i. e. for $\lambda = 0$ the form contributes
   as a dust matter.

   Let us prove relation (\ref{5.d}) for $i \neq j$.
   In what follows we use the following notation
    \beq{5.f}
     C_i ^{~j} =
     \sum_{i_1, \dots, i_p =1}^{n}
      Q_{i i_1 \dots i_p}  Q^{j i_1 \dots i_p}.
     \eeq
  For $i \neq j$, $T[F,g]_i^{~j}$ is proportional to $C_i ^{~j}$.
  This follows from (\ref{2.9i}) and equation (\ref{5.pc}), written in
  components
     \beq{5.pcc}
      F_{u i_1 ... i_{2m}} = e^{2 f(u)} Q_{i_1 ... i_{2m}}.
      \eeq

   Due to (anti-) self-duality of $Q$ we get
     \beq{5.ff}
      C_i ^{~j}
     = \pm \sqrt{|h|}
     \sum_{i_1, \dots, i_p =1}^{n}  \sum_{j_0, \dots, j_p =1}^{n}
      \frac{1}{(p+1)!} \eps _{i i_1 \dots i_p j_0 \dots j_p}
     Q^{j_0 \dots j_p}   Q^{j i_1 \dots i_p} ,
    \eeq
  where $i \ne j$. This can be further   rewritten as
     \bear{5.g}
    C_i ^{~j} &=&  \pm \sqrt{|h|}
    \sum_{i_1, \dots, i_p =1}^{n}  \sum_{j_1, \dots, j_p =1}^{n}
    \frac{1}{p!}\eps _{i i_1 \dots i_p j j_1 \dots j_p}
    Q^{j j_1 \dots j_p}   Q^{j i_1 \dots i_p}
    \nonumber \\ &=& \pm \sqrt{|h|}
    \sum_{i_1, \dots, i_p =1}^{n}  \sum_{j_1, \dots, j_p =1}^{n}
    \frac{(-1)^p}{p!}\eps _{i j_1 \dots j_p  j i_1 \dots i_p}
    Q^{j i_1 \dots i_p}  Q^{j j_1 \dots j_p}
    \nonumber \\
    &=& (-1)^p  C_i ^{~j}.
      \ear
  Note that $j$ is not summed over in the two sums above.
  In going from  (\ref{5.f}) to the first
  line of  (\ref{5.g}) we have carried out $p + 1$ identical
  sums with:  $j_0 = j$,  $j_1 = j$, ..., $j_p = j$, respectively.

  From (\ref{5.g}) one finds that  for odd $p = 2m - 1$
    $$ C_i ^{~j} = - C_i ^{~j} \Rightarrow
                             C_i ^{~j}=0, \quad i \neq j ~$$
  and, hence,  relation (\ref{5.d}) is valid for $i \neq j$.

  Now, we prove  relation (\ref{5.d}) for $i = j$, i.e.
    \beq{5.9}
      T^i_i [F,g] = 0,
    \eeq
  (no summation in $i$) for all $i = 1, \dots, n$.

  It follows from (\ref{2.9i}) and   (\ref{4.pc}), that
    \beq{5.9i}
      T^i_i [F,g] = B(u) [ - \frac{1}{2} \sum_{k =1}^n C^k_k
                           + (q -1) C^i_i ],
    \eeq
  (no summation in $i$) for all $i = 1, \dots, n$, 
  where $B(u)$ is function of $u$.
  The matrix $T^i_j [F,g]$ is trace-less
    \beq{5.10i}
    \sum_{k =1}^n  T^k_k [F,g] = 0,
    \eeq
  since $n = 2 (p + 1) = 2 (q - 1)$. To prove  (\ref{5.9})
  it is sufficient to verify that
    \beq{5.11a}
      T^1_1  = \dots = T^n_n,
    \eeq
  or, equivalently,
    \beq{5.11b}
      C^1_{~1}  = \dots = C^n_{~n}.
    \eeq

   Let us prove without restriction of generality
    $C^1_1  =  C^2_2$. Indeed, using (\ref{5.ff})
    we get (the summation over repeated indices is understood)
     \bear{5.12}
    C_1^{1} &=&  \pm   \frac{\sqrt{|h|}}{(p+1)!}
    \eps _{1 i_1 \dots i_p j_0  \dots j_p}
      Q^{j_0 j_1 \dots j_p}   Q^{1 i_1 \dots i_p}
    \\ \nonumber
      &=&  \pm  \frac{\sqrt{|h|}}{(p+1)!}
      [p \eps _{1 2 i_2 \dots i_p j_0 j_1 \dots j_p}
       Q^{j_0 j_1 \dots j_p}   Q^{1 2 i_2 \dots i_p}
     \\ \nonumber
         &+&  (p + 1) \eps _{1 i_1 \dots i_p  2 j_1 \dots j_p}
                    Q^{ 2 j_1 \dots j_p}   Q^{1 i_1 \dots i_p}]
      \\ \nonumber
        &=&  \pm  \frac{\sqrt{|h|}}{(p+1)!}
        [p \eps _{2 1 i_2 \dots i_p j_0 \dots j_p}
        Q^{j_0 j_1 \dots j_p}   Q^{ 2 1 i_2 \dots i_p}
     \\ \nonumber
              &+&  (p + 1) \eps _{2 j_1 \dots j_p  1 i_1 \dots i_p}
                           Q^{1 i_1 \dots i_p} Q^{ 2 j_1 \dots j_p}]
     \\ \nonumber
          &=&  \pm  \frac{\sqrt{|h|}}{(p+1)!}
          \eps _{2 i_1 \dots i_p j_0 j_1 \dots j_p}
               Q^{j_0 j_1 \dots j_p}   Q^{2 i_1 \dots i_p}
      \\ \nonumber
                     &=&  C_2^{2}.
        \ear

    Here, we used that $p+1 = 2m$ is even.
    Thus,  relations (\ref{5.d}) are proved.

    Now we check the Hilbert-Einstein eqs. (\ref{2.4i}).
    These equations are satisfied for non-diagonal components ($M \neq
    N$), since the Einstein tensor in the  left hand site of (\ref{2.4i})
    for the metric (\ref{5.pa}) is diagonal (see, for example, the
    Appendix in ref. \cite{IMtop}) and stress-energy tensor  (\ref{2.7i})
    $T^M_N$ is also diagonale due to eqs. (\ref{5.d}). For the diagonal
    part of Hilbert-Einstein eqs. (\ref{2.4i}) we obtain the same ordinary
    differential equations (ODE) for  $\phi(u)$ with constant parameter
    $Q^2$ from  (\ref{3.12ca}) as in the flat case. This follows from
    Ricci-flatness of $h$ and (\ref{5.d}).  But the latter ODE were
    checked in our recent paper \cite{DIM}.  It means that we proved that
    relations (\ref{5.pa}),  (\ref{5.pb}) and  (\ref{5.pc}) with
    surrounding notations and assumptions do really define exact solutions
    to field equations in dimensions $D = 4m +1$.

 {\bf Examples of (anti-)self-dual parallel $2m$-forms on
        Ricci-flat Riemannian manifolds of dimension $4m$.}
  It should be noted that such parallel form does exist when
 $4m$-dimensional Riemannian manifold $N$ is K$\ddot{a}$hler Ricci-flat
 manifold with holonomy group $SU(2m)$.
  Indeed the $m$-th wedge power of K$\ddot{a}$hler
  2-form, i.e.
     \beq{e.o}
     \alpha = \Omega^m,
     \eeq
  gives an example of
  non-zero parallel (i.e. covariantly constant) 
  form of rank $2m$ \cite{Besse}.
  Splitting this form into a sum of self-dual and anti-self-dual
  parallel forms:
  \beq{e.1}
  \alpha = \alpha_{+} + \alpha_{-},
  \eeq
  where
   \beq{e.2}
   \alpha_{\pm} = \frac12 (1 \pm *)\alpha
   \eeq
  and * =*[h] is the  Hodge operator on $N$,  we get that either
 $\alpha_{+}$ or $\alpha_{-}$ is a non-zero parallel form. Thus, we get
 an example of either self-dual or anti-self-dual parallel $2m$-form on
 a K$\ddot{a}$hler (Ricci-flat) manifold of dimension $4m$.
 When $N$ is hyper-K$\ddot{a}$hler  Ricci-flat manifold of dimension $4m$
 with holonomy group $Sp(m)$ there are three K$\ddot{a}$hler   2-forms:
 $\Omega_1, \Omega_2, \Omega_3$ .
 In this case we have more  examples of parallel forms
 (e.g. self-dual or anti-self-dual ones),  since any wedge product
   \beq{e.3}
    \alpha = \Omega_1^{m_1} \wedge \Omega_2^{m_2} \wedge \Omega_3^{m_3},
   \eeq
 with $m_1 + m_2 + m_2 = m$, is a parallel 2m-form \cite{Besse}. 
 We should also   mention that there exists a parallel 4-form  on 
 8-dimensional Ricci-flat manifold of  $Spin(7)$ holonomy. 
  See item 10.124 (Table 1) in  \cite{Besse} and also  
  \cite{GP,CGLP} (and references therein).

   \section{Generalization to a set of extra Ricci-flat spaces}

  Here, we suggest a generalization of the solution from the previous section when
  the manifold (\ref{3.10g}) is replaced by

   \beq{3.10gi}
     M = (u_{-}, u_{+})  \times N \times N_1 \times \ldots N_k,
   \eeq
   where $N_r$ are Ricci-flat manifolds with the metric
   $h^r$ of dimension $d_r$, $r = 1, ..., k$.

   The solution reads

     \bear{4A.pai}
   ds^2 =  \exp \left[ \frac{4m f(u)}{K(2-D)} \right]
   \biggl\{ w e^{2c u + 2 \bar c} du^2 \\ \nn
     + \exp (K^{-1} f(u) + 2c^0 u + 2 \bar c^0) h_{ij}(y)dy^i dy^j
     + \sum_{r=1}^{k}  e^{2c^r u+2 \bar{c}^r} ds^2_r \biggr\},
      \\  \label{4A.ppi}
     \varphi^a = - \frac{ \lambda^a}{K} f(u)
     +  c_{\varphi}^a u + \bar c_{\varphi}^a,  \\
     \label{4A.pci}
     F = e^{2 f(u)} du \wedge Q,
         \ear
     $a = 1, \ldots, l$.

     Here $ds^2_r = h^r_{m_r n_r}(z_r) dz_r^{m_r} dz_r^{n_r}$ 
     is the line element
     corresponding to  the metric $h^r$,  $f(u)$ 
     is given by (\ref{4.i}) and
     (\ref{4.ja}), (\ref{4.jb}), (\ref{4.jc}), (\ref{4.jd}),
     \beq{4A.ci}
      c  = 4mc^0 + \sum_{r=1}^k d_r c^r
     \qquad
     \bar c  = 4m \bar c^0 +  \sum_{r=1}^k d_r \bar c^r
     \eeq
     and
     \beq{4A.hbi}
     K  =   \lambda^2 + m + \frac{4m^2}{2-D} \neq 0.
     \eeq

    The integration constants obey the following relations:
    \bear{4A.hai}
     C K^{-1} + m_{ab} c_{\varphi}^a c_{\varphi}^b + 4 m (c^0)^2 \\ \nn
     + \sum_{r = 1}^{k} (c^r)^2 d_r
     - (4 m c^0  + \sum_{r = 1}^{k} c^r d_r)^2 = 0 \\
     2m c^0 = \lambda_a c_{\varphi}^a \qquad
     2m \bar c^0 = \lambda_a \bar c_{\varphi}^a.
     \label{4A.hdi}
      \ear

    The solution is derived in the Appendix.
    More special solutions  were obtained previously in \cite{GIRS,IMS2}.

    When internal spaces $N_1, ..., N_k $ are omitted we get
    the solution from the previous subsection
    with the following identifications between constants:
     \beq{4.NN}
     c_{\varphi}^a  = - \frac{\lambda^a}{K}
     (\lambda_b C_2^b) + C_2^a, \qquad
     \bar c_{\varphi}^a  = - \frac{\lambda^a}{K} (\lambda_b C_1^b) + C_1^a.
     \eeq

  {\bf Solutions with acceleration.}
 Now, we show that among obtained solutions there exist 
 some special solutions with accelerated expansion of certain subspaces.
  Indeed, let us  consider the simplest   cosmological type solution  with
  vanishing integration constants $C = c_{\varphi}^a = c = c^r = 0$.

  This  solution takes place when $K < 0$.
  The metric  (\ref{4A.pai}) for this case  reads
     \beq{5.pai}
       ds^2 = w d \tau^2 +
       B_0 \tau^{2 \nu_0} h_{ij}(y)dy^i dy^j
       + \tau^{2 \nu_1} \sum_{r = 1}^{k} B_r d s^2_r,
     \eeq
  where $\tau > 0$ is "synchronous" variable, $B_0, B_1, \ldots, B_r$
  are positive constants and
     \beq{5.1a}
       \nu_0 = \frac{4m + 2 - D}{2 \Delta},
       \qquad    \nu_1 = \frac{2m}{\Delta}.
     \eeq
 Here
     \beq{5.2a}
       \Delta  = (D-2) K + 2m.
      \eeq

 When $\nu_1 > 1$, or $ - 2m/(D-2) < K < 0$
 we get an accelerated expansion of factor spaces $M_1$, ..., $M_k$.
 This takes place when

     \beq{5.2b}
       -(d+1) m  < \lambda^2  (D-2)  < - (d-1) m.
      \eeq
  where
     \beq{5.lj}
       d  = \sum_{r = 1}^{k} d_r.
      \eeq
  It is possible only if $\lambda^2 < 0$ or, equivalently,
  some of scalar fields are phantom ones.

   We note that solutions of this Section for $\lambda = 0$ are in
   agreement with perfect fluid solutions from \cite{IM95}.

  \section{Examples}

   Here we consider two examples of solutions from the previous
   section.

   \subsection{$II A$ supergravity}

 Let us take the $D=10$ IIA supergravity with the bosonic part of the
 action
 \beq{6.1}
 S=\int d^{10}z\sqrt{|g|}\biggl\{R[g]-(\p\varphi)^2-\sum_{a=2}^4
  \e^{2\lambda_a\varphi}F_a^2\biggr\}-\frac12\int F_4\wedge F_4\wedge A_2,
 \eeq
 where $F_a=dA_{a-1}+\delta_{a4}A_1\wedge F_3$ is an $a$-form and
 \beq{6.2}
  \lambda_3=-2\lambda_4, \quad \lambda_2=3\lambda_4, \quad
 \lambda_4^2=\frac18.
 \eeq

 We consider the solutions with zero forms $A_1$, $A_3$
 (and hence vanishing $F_2$ and $F_4$),
 i.e. we are interested  in the NS-NS sector of the model
 (NS means Neveu-Schwarz).

 Thus, we consider the solution  from the previous section
 describing  for $w = -1$ a ``collection'' of
 electric S1-branes, i.e. S-fundamental strings (SFS).  In this case
 we get $m=1$, $K = 1$ and $k \leq 5$.

 The solution reads
 \bear{4A.paj}
     ds^2 =  \exp(-f(u)/2) \biggl\{ w e^{2c u + 2 \bar c} du^2 \\ \nn
     + \exp (f(u) + 2c^0 u + 2 \bar c^0) h_{ij}(y)dy^i dy^j
     + \sum_{r=1}^{k}  e^{2c^r u+2 \bar{c}^r} ds^2_r \biggr\},
      \\  \label{4A.pbi}
     \varphi = - \lambda_3 f(u) +  c_{\varphi} u + \bar c_{\varphi},  \\
     \label{4A.pcj}
     F_3 = e^{2 f(u)} du \wedge Q.
         \ear

The function $f(u)$   is given by  relation
        \beq{4.ij}
        f = - \ln \left[|z || Q^2|^{1/2} \right]
        \eeq
     with
      \beq{4.jdj}
     z = \frac{1}{\sqrt{C}} \cosh \left[ (u-u_0) \sqrt{C} \right],
        \eeq
     and  the integration constants obey
     \beq{4A.cj}
     c  = 4c^0 + \sum_{r=1}^k d_r c^r
     \qquad  \bar c  = 4\bar c^0 +  \sum_{r=1}^k d_r \bar c^r,
     \eeq
     and
     \bear{4A.haij}
     C = - (c_{\varphi})^2 - 4 (c^0)^2
     -  \sum_{r = 1}^{k} (c^r)^2 d_r  +
     (4 c^0  + \sum_{r = 1}^{k} c^r d_r)^2 > 0 \\
     2 c^0 = \lambda_3 c_{\varphi} \qquad
     2 \bar c^0 = \lambda_3 \bar c_{\varphi}.
     \label{4A.hdij}
      \ear

    For  $w = - 1$  all metrics $h^r$ should be of Euclidean signatures,
    and for $w = + 1$ one metric should be of
    pseudo-Euclidean signature while the other
    ones should be of Euclidean signatures.

   \subsection{Chain of $B_D$-models}

 Now, we consider the chain of the so-called $B_D$-models in dimensions
  $D =11, 12, \ldots$ with a set of "phantom" scalar fields
  suggested in \cite{IMJ} and defined by the action

 \beq{6.14}
  S_D = \int_{M} d^{D}z \sqrt{|g|} \{ {R}[g] +
  g^{MN} \partial_{M} \vec{\varphi} \partial_{N} \vec{\varphi}
  - \sum_{a = 4}^{D-7}
  \frac{1}{a!} \exp( 2 \vec{\lambda}_{a} \vec{\varphi}) (F^a)^2 \},
 \eeq
 where $\vec{\varphi}  = (\varphi^1, \ldots, \varphi^l) \in \R^l$,
  $\vec{\lambda}_a =(\lambda_{a1}, \ldots, \lambda_{al}) \in \R^l$,
  $l = D-11$, ${\rm rank} F^a = a$, $a = 4, \ldots, D-7$.
 Here vectors $\vec{\lambda}_a$ satisfy the relations
  \bear{6.15}
   && \vec{\lambda}_{a} \vec{\lambda}_b = N(a,b)
   - \frac{(a - 1) (b - 1)}{D-2}, \\
   \label{6.16}
   &&N(a,b) = {\rm min}(a,b) - 3,
  \ear
  $a,b = 4, \ldots, D-7$.

The vectors $\vec{\lambda}_a$ are linearly dependent, since
 \beq{6.17}
  \vec{\lambda}_{D-7} = -2 \vec{\lambda}_{4}.
 \eeq
 For $D > 11$  vectors $\vec{\lambda}_{4}, \ldots, \vec{\lambda}_{D-8}$
  are linearly independent.

 The model (\ref{6.14}) contains $l$ scalar fields with negative kinetic
 term (i.e. $m_{a b } = - \delta_{a b}$ in
 (\ref{2.1i})) coupled with $l + 1$ forms. For $D = 11$ ($l= 0$)
 the model (\ref{6.14}) coincides
 with truncated bosonic sector of $D = 11$ supergravity
("truncated" means without Chern-Simons term). For $D = 12$ ($l=1$)
 (\ref{6.14}) coincides with truncated $D = 12$ model from \cite{KKP}
 that is being used for a field description of $F$-theory \cite{Vafa}.

The matrix  (\ref{6.16}) was called in \cite{IMJ} as a fundamental
matrix of the theory since it describes the intersection rules for
branes. The $B_D$-models ("beautiful models" ) have a full set of
binary brane configurations: this means that
$\vec{\lambda}_a$-vectors are chosen in such way that standard
("orthogonal") intersection rule formula \cite{AR,AEH,IMC,IMJ} for
any two brains gives
 us a natural number  for dimension of intersection \cite{IMJ}.

 Let us consider the sector of the $B_D$-model with the form $F_{2m+1}$,
 $m \geq 2$.  It should be $D \geq 2m +8$ in order for this form to
 appear. Now we could write the solution from the previous section
 describing the collection of electric $S(2m -1)$-branes localized on
 $4m$-dimensional sub-manifold $N$. We should also put an obvious
 restriction $D \geq 4m +1$.

   The calculation of $K$-parameter gives us

    \beq{6.K}
    K = 2 - m.
    \eeq

  For $m = 2$ we get $K =0$ and our solution  (from the previous section) 
  does not work, e.g. for $D = 12$ model from \cite{KKP}.  (We are not 
 able also to rewrite our solution for IIB model, since 5-form $F_5$ 
 should be self-dual in this case.)  For $m > 2$ we get $K < 0$ and our 
 solution does work for $D \geq 14$.

    \section{Conclusions and discussions}

  So, we considered a $D =(n+1)$-dimensional cosmological type model
  with several scalar fields and  antisymmetric $(p+2)$-form.
  We  generalized the composite electric $S$-brane solutions from
  \cite{DIM} for $D = 4m+1 = 5, 9, 13, ...$ and   $p = 2m-1
  = 1, 3, 5, ...$ to the case when the form $Q$ of rank $2m$ is defined on
  $4m$-dimensional  oriented Ricci-flat space $N$ of Euclidean signature.
  The Ricci-flat submanifold $N$ may be chosen to be either
  K$\ddot{a}$hler manifold of holonomy group $SU(2m)$,
  or hyper-K$\ddot{a}$hler  manifolds with holonomy group $Sp(m)$,
  or 8-dimensional Ricci-flat manifold of  $Spin(7)$ holonomy.

  Here, the form $Q$ is arbitrary non-zero parallel
  self-dual or anti-self-dual   $2m$-form on $N$.
  For flat $N = \R^{4m}$  the components of this
  form in canonical coordinates are proportional to charge densities of
  electric branes.

  We also  found  generalizations of the solutions to the case when a
  chain of extra Ricci-flat factor-spaces is  added. We have shown that
  these solutions contain as a special case solutions with accelerated
  expansion  of extra factor-spaces.

 We also considered certain examples of solutions:
  for IIA supergravity and a for chain of $B_D$-models in
  dimensions \cite{IMJ} $D = 14, 15, ...$.

 \begin{center}
 {\bf Acknowledgments}
 \end{center}

   The work of V.D.I. and  V.N.M. was supported in part
   by the  DFG grant  Nr. 436 RUS 113/807/0-1 and also
   by the Russian Foundation for
   Basic Researche, grant Nr. 05-02-17478.

  V.D.I. and V.N.M. thanks the colleagues from the Physical Department
  of the University of Konstanz for hospitality during their visits in
  2006.

 We also thank S. Gukov for informing us on certain mathematical
  topics related to this paper.

\renewcommand{\theequation}{\Alph{section}.\arabic{equation}}
\renewcommand{\thesection}{}
\setcounter{section}{0}

 \section{Appendix}

 Here, we present an explicit derivation of the solution from
 Section 5.

 We start with the manifold
 \beq{A.1}
   M = \R  \times M_{0} \times \ldots \times M_{k}
 \eeq
 equipped by the metric
 \beq{A.2}
 g= w \e^{2{\gamma}(u)} du \otimes du +
           \sum_{i=0}^{k} \e^{2\phi^i(u)} h^i,
 \eeq 
 where $w= \pm 1$, $u$ is a distinguished coordinate;
 $h^i$ is a Ricci-flat metric on the  manifold $M_{i}$
 with dimension $d_{i} = \dim M_i$, $i=0,\dots,k$,  $M_0 = N$ is
 oriented manifold, $h^0 = h$ and $d_0 = 4m$.

 We put for potential form:  $A = \Phi(u) \omega $, where
 $\omega$ is a non-zero parallel (anti-) self-dual form on $M_0$
 of rank $2m$ and hence
  \beq{A.3}
   F =  dA =  \dot\Phi (u) du \wedge \omega
  \eeq
  ($\dot x \equiv dx/du$). Here we use identity $d \omega = 0$ for 
  the parallel form $\omega$.                        
  
  We also put

 \beq{A.4}
  \varphi^a= \varphi^a (u),
 \eeq
 $a = 1,...,l$.

The field equations corresponding to the action (\ref{2.1i})
 are equivalent to
equations of motion for the $\sigma$-model with the action
 \beq{A.5}
  S_{\sigma} = \frac{\mu}2
  \int du {\cal N} \biggl\{G_{ij}\dot\phi^i\dot\phi^j
  + m_{a b} \dot\varphi^{a} \dot\varphi^{b}
  + \omega^2 \exp[-2U(\phi,\varphi)] \dot\Phi^2 \},
 \eeq
where
 \beq{A.7}
  {\cal N}= \exp(\gamma_0-\gamma) > 0
 \eeq
is the lapse function with
 \beq{A.6}
 \gamma_0(\phi) \equiv \sum_{i=0}^k d_i \phi^i,
 \eeq

 \beq{A.8}
  U = U(\phi,\varphi)= - \lambda (\varphi) +
  \frac{1}{2} d_0 \phi^0,
 \eeq
  is brane co-vector (linear form) and
 \beq{A.9}
  G_{ij}= d_i\delta_{ij}- d_id_j
 \eeq
 are components of the "pure cosmological" minisupermetric,
 $i,j=0,\dots,k$, see  \cite{IMZ}. We  denote
  \beq{A.omega}
        \omega^2 \equiv \frac{1}{(2m)!}
       h^{i_1 j_1} \dots h^{i_{2m} j_{2m}}
       \omega_{i_1  \dots i_{2m}} \omega_{j_1  \dots j_{2m}} > 0,
        \eeq
  the square of the non-zero $\omega$-form on the
  Riemannian manifold $M_0$.

For finite space volumes $V_i$ of $M_i$ the action (\ref{2.1i})
just coincides (up to a surface term irrelevant for classical
consideration) with the action (\ref{A.5}) if $\mu=-w V_0 \ldots
V_n$. The non-diagonal part of Hilbert-Einstein equations is
satisfied due to (anti-) self-duality condition on $\omega$: this
may be verified as it was done in section 4. Diagonal part of
Hilbert-Einstein equations and other field equations are governed
by the action (\ref{A.5}).

Action (\ref{A.5}) may be also written in a more condensed form
 \beq{A.10}
  S_\sigma=\frac{\mu}{2}
  \int du {\cal N} \biggl\{ \bar G_{AB}\dot x^A\dot x^B
  + \omega^2 \exp[-2U(x)] \},
 \eeq
 where $x = (x^{A})=(\phi^i,\varphi^a)$,

 \bear{A.15}
       (\bar G_{AB})=\left(\begin{array}{cc}
                                   G_{ij}&0\\
                                   0&m_{a b}
 \end{array}\right),
\ear
  $U(x) =U_A x^A$ is defined in  (\ref{A.8}) and
 \beq{A.17}
  (U_A )=(\frac12 d_0 \delta_{i}^0, - \lambda_a).
 \eeq

 We put
 \beq{A.18}
 (U,U)=\bar G^{AB} U_A U_B \neq 0,
 \eeq
where
 \beq{A.19}
 (\bar G^{AB})=\left(\begin{array}{cc}
  G^{ij}&0\\
   0&m^{a b}
 \end{array}\right)
 \eeq
is the matrix inverse to (\ref{A.15}).
 Here (as in \cite{IMZ})
 \beq{A.20}
   G^{ij}=\frac{\delta^{ij}}{d_i}+\frac{1}{2-D},
 \eeq
 $i,j=0,\dots,k$. The scalar product
 (\ref{A.18}) is
 \beq{A.21}
 (U,U)= K,
 \eeq
 with $K$ defined in Section 5.

 In what follows, we will use contra-variant components
                  $U^A = \bar G^{AB}  U_B$ :
 \beq{A.22}
  U^{i}= G^{ij}U_j= \frac{1}{2} (\delta_{i}^0 -\frac{d_0}{D-2}), \quad
  U^{a}= -  \lambda^a.
 \eeq

The problem of integrability may be simplified if we integrate the
generalized Maxwell  equations \bear{A.23}
 \frac d{du}\left(\exp(-2U) \dot \Phi \right)=0
 \Longleftrightarrow
 \dot\Phi= q \exp(2U),
\ear where $q \neq 0$ is a constant (we are not interested in the
trivial case $q =0$).

Let  us denote
 \beq{A.Q}
  Q = q \omega
  \eeq
and fix the time-gauge to be harmonic one, i.e. we put
 \beq{A.h}
       \gamma = \gamma_0.
  \eeq
 Obviously, $Q^2 > 0$.

Then, the Lagrange equations for the action
 (\ref{A.10}) corresponding to $(x^A)=(\phi^i,\varphi^a)$,
when equations (\ref{A.23}) are substituted, are equivalent to the
Lagrange equations for the Lagrangian
 \beq{A.24}
 L_Q=\frac12 \bar G_{AB}\dot x^A\dot x^B-V_Q,
 \eeq
 with the  zero-energy constraint  imposed
 \beq{A.26}
 E_Q= \frac12 \bar G_{AB} \dot x^A\dot x^B+V_Q=0.
 \eeq
 Here,
 \beq{A.25}
  V_Q=  \frac12 Q^2 \exp[2U(x)].
 \eeq

Then, the Euler-Lagrange equations for the Lagrangian (\ref{A.1})
have the following general solutions \cite{GIM}
 \beq{A.35}
  x(u)= - \frac{U}{(U,U)} \ln | \bar f (u-u_{0})|
       + c u + \bar c,
 \eeq
where $c$ and $\bar c$ are constant vectors obeying,
 \beq{A.36}
  U(c)= U (\bar c)=0 ,
 \eeq
and 
\bear{A.37}
 \bar f (\tau)= 
 \left|\frac{Q^2}{2E_1}\right|^{1/2}\sinh(\sqrt{C}\tau),
 \; C>0, \;   (U,U)<0 , \\  \label{A.38}
 \left|\frac{Q^2}{2E_1}\right|^{1/2}\sin(\sqrt{|C|}\tau),
 \; C<0, \;  (U,U)<0 , \\  \label{A.39}
 \left|\frac{Q^2}{2E_1} \right|^{1/2}\cosh(\sqrt{C}\tau),
 \; C>0, \;  (U,U)>0 , \\  \label{A.40}
 \left|Q^2 (U,U) \right|^{1/2} \tau,
 \; C = 0, \;  (U,U) <0 ,
 \ear
 $C= 2 E_1 (U,U)$, $E_1$, $u_{0}$ are constants.

  For the energy corresponding to the solution  we have
  \beq{A.41}
   E_Q= E_1+ \frac12 (c,c).
  \eeq

 Rewriting this solution in components
  $x = (x^{A})=(\phi^i,\varphi^a)$ and using the relations
 presented above and $f = - \ln| \bar f|$
 we are led to the solution from Section 5.


  \end{document}